\documentclass[prx,twocolumn,showpacs,superscriptaddress]{revtex4-2}
\usepackage[T1]{fontenc}      
\usepackage{epsfig}
\usepackage{amsmath,amssymb}
\usepackage{mathtools}
\usepackage{graphicx}
\usepackage{footnote}
\usepackage{bbm}
\usepackage{bm}

\usepackage[normalem]{ulem}
\usepackage{color}
\definecolor{orange}{rgb}{1, 0.5, 0}
\definecolor{darkgreen}{rgb}{0, 0.5, 0}

\usepackage[all]{xy}

\newcommand{\ceq}[1]{Eq.~(\ref{#1})}
\newcommand{\cfg}[1]{Fig.~\ref{#1}}

\newcommand{\etal}{\emph{et al.}}
\newcommand{\dga}{D$\Gamma$A}
\newcommand{\sqsq}{\ensuremath{\sqrt{3}\!\times\!\sqrt{3}}}
\newcommand{\qzero}{\ensuremath{\mathbf{q}\!=\!0}}
\newcommand{\vbz}{\ensuremath{{V_{\!\text{\tiny BZ}}}}}
\newcommand{\ubz}{\ensuremath{{\!\text{\tiny BZ}}}}

\definecolor{niceblue}{RGB}{0,50,150}
\definecolor{nicered}{RGB}{200,50,0}

\usepackage[colorlinks,breaklinks,bookmarks=true,citecolor=niceblue,linkcolor=nicered,urlcolor=niceblue]{hyperref}

\begin{document}

\title{How correlations change the magnetic structure factor of the kagome Hubbard model}

\author{Josef~Kaufmann}
\affiliation{Institute for Solid State Physics, TU Wien, 1040 Vienna, Austria}
\affiliation{Institute for Theoretical Solid State Physics, IFW Dresden, 01069 Dresden, Germany}

\author{Klaus~Steiner}
\affiliation{Institute for Solid State Physics, TU Wien, 1040 Vienna, Austria}
\affiliation{Department of Physics, University of California, Davis, California 95616, USA}

\author{Richard~T.~Scalettar}
\affiliation{Department of Physics, University of California, Davis, California 95616, USA}

\author{Karsten~Held}
\affiliation{Institute for Solid State Physics, TU Wien, 1040 Vienna, Austria}

\author{Oleg~Janson}
\affiliation{Institute for Theoretical Solid State Physics, IFW Dresden, 01069 Dresden, Germany}

\date{\small\today}
\begin{abstract}
The kagome Hubbard model (KHM) is a paradigmatic example of a frustrated two-dimensional model. 
While its strongly correlated regime, described by a Heisenberg model, 
is of topical interest due to its enigmatic prospective spin-liquid ground state, 
the weakly and moderately correlated regimes remain largely unexplored. 
Motivated by the rapidly growing number of metallic kagome materials 
(e.g., Mn$_3$Sn, Fe$_3$Sn$_2$, FeSn, Co$_3$Sn$_2$S$_2$, Gd$_3$Ru$_4$Al$_{12}$), 
we study the respective regimes of the KHM by means of three complementary numerical methods:
the dynamical mean-field theory (DMFT), the dynamical vertex approximation (D$\Gamma$A),
and determinant quantum Monte Carlo (DQMC). In contrast to the archetypal square-lattice, 
we find no tendencies towards magnetic ordering, as magnetic correlations remain short-range. 
Nevertheless, the magnetic correlations undergo
a remarkable crossover as the system approaches the metal-to-insulator transition.
The Mott transition itself does however not affect the magnetic correlations.
Our equal-time and dynamical structure factors can be used as a reference
for inelastic neutron scattering experiments on the growing family of metallic kagome materials.
\end{abstract}
\maketitle

\section{Introduction} \label{sec:Intro}
Electronic correlations are at the origin of a large  variety of exotic phases
in transition metal oxides and intermetallic compounds~\cite{dagotto05,
pesin10, basov11}.  Yet, an adequate description of correlated materials
remains a long-standing conundrum in condensed matter physics~\cite{imada10}.
Pioneering works of Kanamori~\cite{kanamori63}, Hubbard~\cite{hubbard63}, and
Gutzwiller~\cite{gutzwiller63} enlightened the key role of onsite Coulomb
repulsion as the leading source of electronic correlations. The ensuing model,
known as the Hubbard model, became a paradigm of correlated materials: it is
widely used to rationalize the experimentally observed phenomena such as
metal-to-insulator transitions (MIT)~\cite{imada98}, the formation of local
moments~\cite{anderson61} and high-temperature
superconductivity~\cite{pickett89}.

\begin{figure}[tb]
  \centering
  \includegraphics[width=8.6cm]{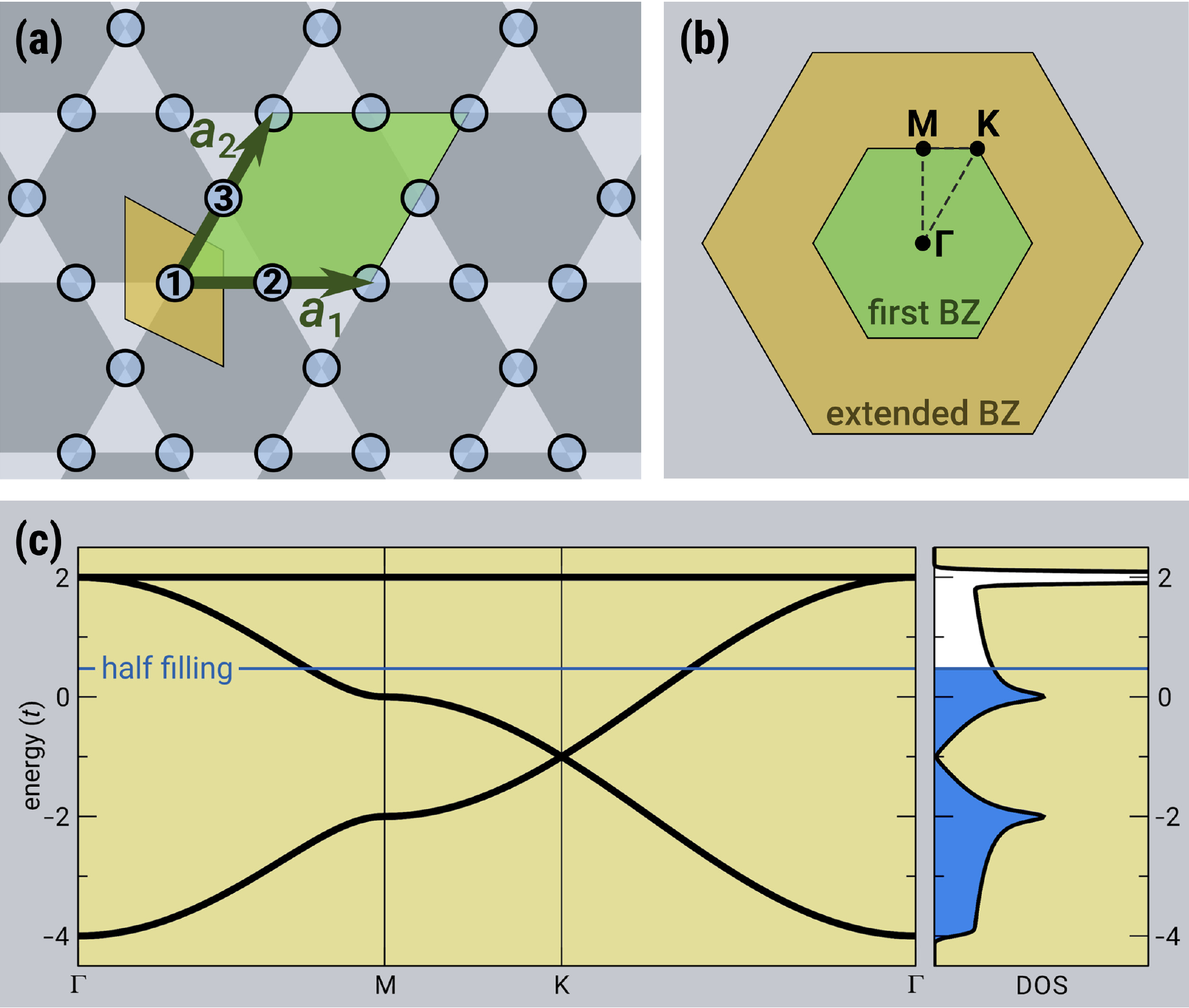}
  \caption{\label{fig:overview} (a) Unit cell (shaded green) of the kagome
lattice described by the basis vectors $a_1$ and $a_2$ comprises three sites
labeled 1, 2, and 3. The Wigner-Seitz cell is shaded ochre. (b) High-symmetry
points of the reciprocal space, the first and the extended Brillouin zones
(BZ). (c) The band structure (left) and the density of states (DOS, right) of
the tight-binding kagome model with a positive $t$. The blue line marks the
chemical potential $\mu_\text{tb}\approx0.47$, for which the non-interacting
system is half-filled.
}
\end{figure}

The flip side of such a rich physics and complexity is that a solution of the Hubbard model
becomes intractable for physically relevant regimes.  Two prominent exceptions
are the one-dimensional~\cite{lieb68} and the infinitely-dimensional Hubbard
models.  The latter can be solved within the dynamical mean-field theory
(DMFT)~\cite{georges92, georges96}, a self-consistent mapping of a lattice
Hubbard model onto a single-site Anderson impurity model.  This mapping is
justified by the rapid suppression of nonlocal correlations concomitant with
an increase of the spatial dimensionality or the nearest neighbor coordination number~\cite{metzner89, mueller-hartmann89}.
While in infinite dimensions the DMFT offers an exact solution of the Hubbard
model~\cite{jarell92}, it remains a well-justified approximation for realistic
three-dimensional models. 

The restriction to local correlations may break down in two-dimensional models
relevant for various experimental situations, e.g.\ in high-temperature
superconductors~\cite{keimer15}, ultrathin van der Waals
superlattices~\cite{cao18}, and oxide heterostructures~\cite{mannhart10,
hwang12}. Coordination numbers in such systems are typically small, and hence the outcome of an onsite scattering event strongly depends on the configuration of the neighboring sites. 
A striking example showing the importance of nonlocal correlations is the
square-lattice Hubbard model at half-filling, which describes the physics of
parent compounds of cuprate high-temperature superconductors. Here, DMFT finds
a first-order Mott transition from a metal to a paramagnetic insulator at low
temperatures.  Nonlocal correlations drastically change the physics: while
short-range antiferromagnetic correlations significantly reduce the critical
interaction $U$, long-range antiferromagnetic correlations shift the MIT to much smaller interactions, even to $U=0$ for a square lattice Hubbard model with perfect nesting~\cite{schaefer15, schaefer20}.  Long-range antiferromagnetic correlations open a gap.
This conclusion is corroborated by quantum Monte Carlo simulations on large
finite lattices~\cite{white89,hirsch89,schaefer15, varney09}. For two-dimensional lattices without perfect nesting these results suggest a MIT at still considerably lower values of $U$. As we will see below, for a magnetically frustrated model with only short-range antiferromagnetic correlations, such as the KHM, DMFT only slightly overestimates the critical $U$ of the MIT.

The situation becomes qualitatively different in the presence of frustration.
As sizable ground-state degeneracy is typical for a frustrated model, magnetic
instabilities may spread over different momenta, suppressing the overall
tendency towards magnetic ordering. Ramifications for the MIT are surmised, but
model-specific information is scarce. The most studied case, the Hubbard model
on the triangular lattice, shows a conventional MIT even if nonlocal
correlations are taken into account. However, the triangular lattice has
the sizable coordination number of six bonds per site, which is the same as in
the 3D cubic lattice. As a result, quantum fluctuations in the triangular lattice 
are largely suppressed \cite{Antipov2011,Li2014}.

The kagome lattice (Fig.~\ref{fig:overview}, a) is a much more apt playground
to study the interplay of nonlocal correlations and frustration.  This simple
tessellation of regular hexagons and triangles exhibits a remarkably involved
magnetism: the ground state of the $S=\frac12$ Heisenberg model on a kagome
lattice remains under debate for several decades, with the handful of candidate
states comprising a Dirac spin liquid \cite{ran07, hermele08, ma08, iqbal13,
he17, liao17, chen18}, a $Z_2$ spin liquid \cite{yan11, depenbrock12, jiang12,
nishimoto13, kolley15, mei17, laeuchli19}, a chiral spin liquid \cite{messio12,
capponi13a, gong14, bauer14, he14, wietek15, gong15, hu15, messio17}, and a
valence bond solid \cite{marston91, syromyatnikov02, nikolic03, singh07,
budnik04, evenbly10, schwandt10, poilblanc10, poilblanc11}.  The vibrant
research is boosted by experimental studies of herbertsmithite
$\gamma$-Cu$_3$Zn(OH)$_6$Cl$_2$~\cite{herb:helton07, han12, khuntia20}. This
correlated insulator material does not show any sign of magnetic ordering down
to lowest temperatures~\cite{mendels07}, despite the sizable antiferromagnetic
exchange of $\sim$200\,K~\cite{janson12, jeschke13}.  As typical for cuprates,
the (screened) onsite Coulomb repulsion $U$ largely exceeds the intersite
hopping amplitude $t$, placing herbertsmithite into the strongly correlated
limit of the Hubbard model ($U\gg{}t$). Interactions between localized
$S$\,=\,$\frac12$ spins in this limit are well-described by the Heisenberg
model.

Much less is known about less correlated regimes of the half-filled
kagome Hubbard model (KHM).  Dynamical spin correlations were studied using
cluster DMFT on a $N$\,=\,3 site clusters~\cite{ohashi06, ohashi07} where the
first-order metal to insulator transition (MIT) occurs at
$U_{\text{c}}=8.4t$~\cite{ohashi06}. As was noticed in Ref.~\cite{higa16}, a
major drawback of odd-numbered clusters is their incompatibility with
valence bond states. Instead, variational cluster calculations on $N$\,=\,6 and
$N$\,=\,12 clusters demonstrate that the formation of intersite singlets
(valence bonds) underlies the MIT, and propels it to a smaller $U$ value.  A
very recent study suggests that the KHM harbors a topologically nontrivial state, a
higher-order topological Mott insulator, characterized by corner modes whose
spin excitation spectrum is gapless~\cite{kudo19}.

For a long time, the KHM attracted little attention: although their potential
modifications of herbertsmithite hold promise for unconventional
phases~\cite{mazin14}, pristine herbertsmithite and related quantum magnets~\cite{janson08, fak12, iqbal15} are
correlated insulators deep in the Heisenberg limit, The situation changed
drastically after the discovery of metallic kagome materials
Mn$_3$Sn~\cite{nakatsuji15, nayak16, kuroda17, kimata19, li19c, wuttke19},
Fe$_3$Sn$_2$~\cite{fenner09, kida11, hou17, ye18, yin18, lin18, li19a, li19b,
tanaka20}, Co$_3$Sn$_2$S$_2$~\cite{liu18, wang18, yin19, liu19, shen19,
lachman20, howard19X, xing19X, yin20, liu21},
Gd$_3$Ru$_4$Al$_{12}$~\cite{nakamura18, matsumura19}, and very recently,
FeSn~\cite{inoue19, kang19, lin20, sales19}.  Interestingly, the first three
materials exhibit a sizable anomalous Hall effect, associated with a
nonvanishing Berry curvature of the occupied bands~\cite{nayak16, ye18, liu18}.
While the magnetic moments are associated with the electrons localized in $3d$
or $4f$ shells, neither of the five materials is insulating.  Hence, a key to
their electronic and magnetic properties should be sought in less correlated
regimes of the KHM, which remain hitherto largely unexplored.

In this paper, we fill this gap by performing an extensive numerical
investigation of the KHM using three different many-body techniques: the
determinant quantum Monte Carlo (DQMC)\cite{blankenbecler81}, the
dynamical mean-field theory (DMFT)~\cite{georges92, georges96},  and the
dynamical vertex approximation (\dga)~\cite{toschi07,rohringer18}. DQMC
is a numerically exact technique for fermionic lattice models. With the caveat
that finite lattices beget finite size effects, it provides a sound benchmark
for quantum impurity methods.  Since frustration of the KHM gives rise to a
severe sign problem, we restrict our DQMC calculations to relatively high
temperatures.  We use these results as a benchmark for \dga, a diagrammatic
extension of DMFT. In contrast to cluster extensions of DMFT~\cite{maier05},
this method accounts for nonlocal correlations on all length scales --- from
short-range to long-range. And unlike many QMC-based techniques, diagrammatic
extensions of DMFT are immune to the sign problem~\cite{rohringer18}, allowing
us to explore more correlated regimes of the KHM. In this study, we apply the
recently implemented self-consistent \dga\ scheme~\cite{kaufmann20X}, which
eliminates the need to restore the sum rules by means of so-called
$\lambda$-corrections \cite{Katanin2009,Rohringer2016}.

Our main finding is the gradual correlation-induced change in the regime of
magnetic correlations: While maxima at the $K$-point of the extended Brillouin
zone are indicative of dominant \sqsq\ correlations, the enhancement of
interaction strength gives rise to the sign change of third-neighbor
correlations. Interestingly, this crossover occurs in the metallic phase, while
spin correlations in the moderately correlated regime are similar to those of
the Heisenberg model. This finding gives us a key to distinguish between weakly
and strongly correlated regimes in the growing family of kagome materials.
Furthermore, we compute the dynamical structure factors $S(\mathbf{q},\omega)$
for the different regimes of the kagome Hubbard model. Since these quantities
are accessible in inelastic neutron scattering  experiments, the relative
strength of electronic correlations in real materials can be estimated by a
direct comparison to our calculated $S(\mathbf{q},\omega)$.

Our study is equally important for method development in the field of
electronic correlations: it applies a diagrammatic beyond-DMFT method, \dga, to
a strongly frustrated two-dimensional model. Extensive comparisons with the
numerically exact lattice-based method (DQMC) reveal an overall good agreement,
indicating that a self-consistent \dga\ calculation captures the leading
effects of nonlocal fluctuations, even if tendencies towards magnetic ordering
are strongly suppressed.

This paper is organized as follows. In Sec.\ \ref{sec:model-methods} we
introduce the Hubbard model on a kagome lattice, and briefly
explain the methods we use to obtain our results. The main results are
presented in Sec.\ \ref{sec:results}, where we first present the phase diagram
and then discuss the magnetic structure factors. In Sec.\ \ref{sec:discussion}
our results are compared to previous theoretical results from the literature
and put in the context of present-day experimental research. We summarize our
results in Sec.~\ref{sec:conclusion}.
Additionally we provide more detailed information about the influence
of certain real-space correlations on the structure factor in Appendix \ref{app:corr}. 

\section{Numerical methods}\label{sec:model-methods}
\subsection{KHM Hamiltonian}
We define the Hamiltonian of the Hubbard model on a kagome lattice as
\begin{align}
  \label{eq:HM}
  H = \frac{1}{\vbz} \int_\ubz d\mathbf{k}\sum_{jl,\sigma} 
      &h_{jl}(\mathbf{k}) c_{j\sigma}^\dag(\mathbf{k}) c_{l\sigma}(\mathbf{k})\notag \\
      +& \sum_\mathbf{R} \sum_j U n_{\mathbf{R}j\uparrow} n_{\mathbf{R}j\downarrow},
\end{align}
where $\vbz$ is the volume of the Brillouin zone (\cfg{fig:overview}, b),
$\mathbf{k}$ is the 2D crystal momentum, and indices $j$ and $l$ refer to the sites
within the unit cell and run from 1 to 3. The three sites comprising the unit cell
(as shown in \cfg{fig:overview}, a) form an equilateral triangle whose side length is a half of the lattice constant.  The tight-binding Hamiltonian $h_{jl}(\mathbf{k})$ incorporates the lattice geometry and hopping amplitudes.  In the second term, we have a sum over all unit cells, where site $j=1$ is located at the Bravais lattice position $\mathbf{R}$. The interaction, parametrized by a scalar $U$, is of density-density type.

As hopping is allowed only between neighbor sites, the non-interacting part of
the Hamiltonian reads:
\begin{equation}
  \label{eq:dispersion-matrix}
  h_{jl}(\mathbf{k}) = -t \left(
  \begin{matrix}
    0 & 1+e^{ik_1} & 1+e^{ik_2}\\
    1+e^{-ik_1} & 0 & 1+e^{-i(k_1-k_2)}\\
    1+e^{-ik_2} & 1+e^{i(k_1-k_2)} & 0
  \end{matrix}
  \right),
\end{equation}
where $k_1$ and $k_2$ are the projections of $\mathbf{k}$
onto the reciprocal basis vectors $\mathbf{b}_1$ and $\mathbf{b}_2$.
We set the hopping amplitude to unity $t\equiv 1$, which defines our unit of
energy used throughout the paper. By further setting $\hbar\equiv 1$ and
$k_B\equiv 1$ we also fix the units of frequency and temperature.

By applying a unitary transformation $\mathcal{U}(\mathbf{K})$, 
the Hamiltonian matrix \ceq{eq:dispersion-matrix}
can be diagonalized
\begin{equation}
  \label{eq:tb-diagonalization}
  h(\mathbf{k}) = \mathcal{U}(\mathbf{k})\; \varepsilon(\mathbf{k})\!\mathbbm{1}\; \mathcal{U}^\dag(\mathbf{k}),
\end{equation}
where $h$ and $\mathcal{U}$ are $\mathbf{k}$-dependent matrices of dimension
three,
$\mathbbm{1}$ is a 3$\times$3 identity matrix, and $\varepsilon$ is a
$\mathbf{k}$-dependent three-dimensional vector, defining the tight-binding
bands. 
The latter are shown in \cfg{fig:overview}, c.
Remarkably, one band is completely flat, corresponding to a $\delta$-peak in the
density of states (DOS). The two dispersive bands are identical to those in a honeycomb lattice, featuring two Van Hove singularities and a Dirac crossing. The eigenstates of the flat band correspond to states that are localized on hexagonal plaquettes and combinations thereof \cite{bergman08}.

Evaluation of eigenenergies of the full Hubbard Hamiltonian \ceq{eq:HM} is not possible:
the tight-binding term and the interaction term do not commute. 
We are therefore restricted to a handful of numerical methods that 
allow us to calculate correlation functions within a certain approximation. 
In this work, we use three many-body methods to compute properties of the Hubbard model 
on a kagome lattice: the dynamical mean-field theory (DMFT), 
the dynamical vertex approximation (\dga), and the determinant quantum Monte Carlo (DQMC).
Since the kagome lattice is not a standard application of these methods, 
we briefly review how they work in this case in order to prevent confusion.

\subsection{Dynamical mean-field theory} 
The dynamical mean-field theory (DMFT) utilizes the equivalence of the Hubbard model in infinite dimensions to an Anderson impurity model. The latter is amenable to an exact numerical evaluation of correlation functions. The pertinent hybridization function of the (auxiliary) Anderson impurity model is determined self-consistently \cite{georges96}. While DMFT self-energies are frequency-dependent, they lack momentum dependence, i.e., they are local.

Although DMFT workflows are exhaustively described in the literature, we nevertheless provide an outline of our calculation scheme for the sake of clarity. 
In each step of a DMFT calculation, the four following operations are performed:

\begin{enumerate}
  \item Calculate the local Green's function $G_\text{loc}(i\omega_n)$ for the Hubbard model:
    $$
      \quad\quad\;
      G_\text{loc}(i\omega_n) = \frac{1}{\vbz} \!\!\int_\ubz\!\!\!\!\! d\mathbf{k}
        \Big[\! \big(i\omega_n \!\!+\! \mu \!-\! \Sigma(i\omega_n)\big)\mathbbm{1} - h(\mathbf{k}) \Big]^{-1}
    $$
    At this step the chemical potential $\mu$ is adapted in order to keep the system half filled.
    Note that the local Green's function $G_\text{loc}(i\omega_n)$  at fermionic Matsubara frequency $\omega_n$ 
    is a still a matrix with respect to the three $j$ sites at  Bravais lattice site ${\mathbf{R}=0}$, 
    whereas the self-energy $\Sigma(i\omega_n)$ is scalar multiplied with the $3\times 3$ unit matrix $\mathbbm{1}$.
  \item Calculate the non-interacting Green's function $\mathcal{G}_j(i\omega_n)$ for the impurity model at site $j$ within the unit cell:
    $$
      \left[{\mathcal{G}_j(i\omega_n)}\right]^{-1} = \biggl[{\bigl[G_\text{loc}(i\omega_n)\bigr]}_{jj}\biggr]^{-1} + \Sigma(i\omega_n)
    $$
    This  ${\mathcal{G}_j(i\omega_n)}$ defines three (equivalent) impurity problems for each site $j$ of the unit cell,  consistent with the DMFT approximation. 
  \item Calculate the self-energy $\Sigma(i\omega_n)$ of the impurity models defined in step 2 by an impurity solver. To this end, we use the numerically exact continuous-time quantum Monte Carlo (CT-QMC) algorithm in the hybridization expansion~\cite{gull2011}. Since all three impurities are equivalent, we need to do this calculation only once, thus saving computational time.
  \item Insert the self-energy $\Sigma(i\omega_n)$ of step 3 into the expression for the Green's function $G_\text{loc}(i\omega_n)$ in step 1 and iterate until convergence.
\end{enumerate}

Our calculations were carried out by the program package w2dynamics \cite{w2dynamics}. Although we are solving a three-band model, the downfolding approximation reduces the numerical cost of the DMFT calculation roughly to that of a one-band calculation. By using symmetric improved estimators \cite{kaufmann19}, our DMFT calculations converge very precisely and the resulting self-energies are practically free of noise. Using the final impurity model of the converged DMFT calculation, we compute also the generalized susceptibility in CT-QMC by worm sampling \cite{gunacker15}.

\subsection{Dynamical vertex approximation (D$\mathbf{\Gamma}$A)}

The \dga\ \cite{toschi07,rohringer18} is a method based on Feynman diagrams.
Compared to DMFT, where the self-energy is local, \dga\ goes one step further
and imposes locality on the irreducible two-particle vertex.  The
Bethe-Salpeter equation combines these local building blocks by nonlocal
propagator lines and leads to a momentum dependence in the susceptibility.  A
momentum-dependent self-energy is then obtained via the Schwinger-Dyson
equation of motion.  In our calculations we use ladder-\dga, where nonlocal
fluctuations are considered only in the density (charge) and the magnetic (spin)
channel.  In the following, we sketch the \dga\ procedure in a compact tensor notation suppressing the frequency and for the lattice quantities momentum and three basis site indices: the interested
reader will find a more detailed description in \cite{galler17} and
particularly in \cite{kaufmann20X}.

The main input of a \dga\ calculation is, besides the tight-binding Hamiltonian,
the irreducible vertex $\Gamma_d$($\Gamma_m$) in the density (magnetic) channel.
It is computed from the generalized susceptibility $\chi$ of the DMFT impurity by the Bethe-Salpeter equation
\begin{equation}
  \label{eq:bse-imp}
  \Gamma_r = \beta^2 \big(\chi_{r}^{-1} - \chi_0^{-1}\big),
\end{equation}
where $r$ denotes the channel.
Note that the inversion is done only with respect to fermionic Matsubara frequencies, since the impurity problem has no orbital degrees of freedom. Next, we perform the following iterative procedure \cite{kaufmann20X}:
\begin{enumerate}
  \item Calculate momentum-dependent reducible vertices $F_d$ and $F_m$ for the kagome lattice  by the Bethe-Salpeter equation
    \begin{equation}
      \label{eq:bse-latt}
      F_{r} = \big[\mathbbm{1} - \frac{1}{\beta}\Gamma_{r}GG\big]^{-1}\Gamma_{r}\text{ with }r=d\text{ or }m.
    \end{equation}
    Here, the inversions pertain also to the site indices $j$ and $l$, since $G$ is the Green's function of the Hubbard model and thus a 3$\times$3 matrix.
  \item Combine $F_d$ and $F_m$ to a crossing symmetric vertex 
    $\mathbf{F}$ as explained in Ref.~\cite{galler17}.
  \item Compute the momentum-dependent self-energy by the equation of motion, 
	  which schematically reads
    \begin{equation}
      \label{eq:eom}
	    \Sigma(\mathbf{k}) = \frac12 Un + \frac{1}{\beta^2}U GG \mathbf{F} G.
    \end{equation}
    The first term is the static (Hartree) contribution, the second one contains the diagrams of higher order.
  \item Construct a new lattice Green's function $G$ by
    \begin{equation}
      \label{eq:dyson-lattice}
      \hspace{2em}
      G(\mathbf{k}, i\omega_n)
       = \Big[\big(i\omega_n \!+\! \mu)\mathbbm{1} 
         - h(\mathbf{k}) - \Sigma(\mathbf{k}, i\omega_n) \Big]^{-1}
    \end{equation}
    Similar to DMFT, the chemical potential $\mu$ can be (slightly)
    adapted in order to keep the system half filled.
    This Green's function is now used as a propagator in step 1
    and the steps are repeated until convergence in $\Sigma$.
\end{enumerate}

After convergence is reached, we obtain the self-energy 
$\Sigma(\mathbf{k},i\omega_n)$, which is a full 3$\times$3 matrix in the space of the three lattice sites of our basis and depends, in addition to the fermionic Matsubara frequency, also on the crystal momentum $\mathbf{k}$.
A similar procedure for unit cells with multiple equivalent one-orbital impurities
was also proposed for the dual fermion approach \cite{hirschmeier18}.

\subsection{Determinant quantum Monte Carlo (DQMC)}
Mapping onto an auxiliary impurity problem lies at the core of both DMFT and
\dga\ approaches. It is therefore crucial to cross-check the DMFT/\dga\ results
with an independent numerically exact method working explicitly with a
\emph{lattice} problem. To this end, we employ determinant quantum Monte Carlo
(DQMC) simulations \cite{blankenbecler81}. This technique explicitly computes nonlocal correlation
functions, such as the lattice Green's function. Therefore, the main
approximation lies in the restriction to a finite cluster.
As a consequence, quantities such as the susceptibility or equal time structure
factor, which sample correlations to large distance, have corrections which
scale as $1/L$, where $L$ is the linear lattice size.  However, local
(e.g.~near neighbor) correlation functions are usually converged to a few
percent on lattices of linear extent $L\sim 10$. 
In the present case computations were done for clusters of 5$\times$5 unit cells
(i.\ e.\ 75 sites).
DQMC calculations also have
`Trotter' errors proportional to the square of the discretization interval of
the inverse temperature $\beta=1/T$.  In our work Trotter errors are of the same
order as, or smaller than, the statistical errors from the Monte Carlo
sampling.  Finally, as noted earlier, DQMC calculations are limited by the sign
problem~\cite{loh90,iglovikov15}.  A rough rule of thumb is that DQMC can be
done down to temperatures $T \sim W/30$, where $W$ is the bandwidth, for
interaction strengths $U \sim W$.

\section{Results}\label{sec:results}
We divide our results in two parts. In Sec.~\ref{sec:phasediag}, we focus on the phase diagram of the KHM in the metallic regime and discuss how the spectral function evolves as a function of the interaction strength and temperature. While 
it is a common way to describe the physics of a model, we surmise that from the experimental viewpoint such a discussion is largely disconnected from the physics of kagome materials: there is no recipe to extract the information on the interaction strength $U$ for a given material. To alleviate this problem, we present our results for the equal-time $S(\mathbf{q})$ as well as dynamical $S(\mathbf{q}, \omega)$ structure factors for different $U$ values in Sec.~\ref{sec:sq}. As these quantities are experimentally accessible, a direct comparison between theory and experiment can yield an estimate for $U$.

\subsection{Phase diagram of the kagome Hubbard model}\label{sec:phasediag}
The Hubbard model is famous for exhibiting a transition between a metallic state and an insulating state upon variation of the interaction strength. This generic trend holds for the kagome lattice:
In the KHM, the local spectral functions
\footnote{All analytic continuations are done using the maximum entropy method as implemented in
ana-cont~\cite{kaufmannGithub}.} (\cfg{fig:spec-loc}) show a reduction of the
spectral weight around the Fermi level (at $\omega=0$) upon increasing $U$, and
the system eventually turns insulating once the interaction exceeds a certain
critical value $U_c$.  In the high-temperature regime ($T=0.33$), some spectral
weight remains in the gap, but the spectral function develops a distinct dip at
the Fermi level, with $U_c\approx 7$ in DQMC and $U_c\approx 9$ in DMFT.

\begin{figure}[tb]
  \includegraphics[width=0.5\textwidth]{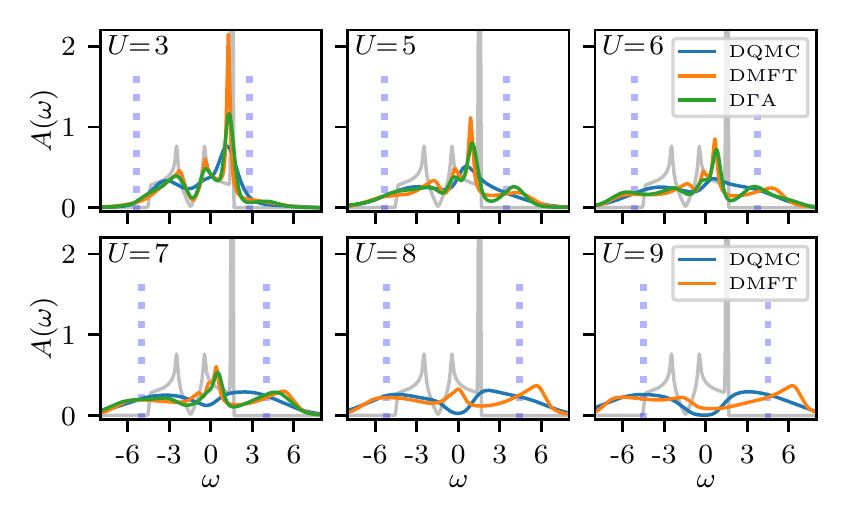}
  \caption{\label{fig:spec-loc}Local spectral functions for interaction strength
    ranging from  $U=3$ to 9 at $T=0.33$ in units of the hopping $t\equiv 1$ as obtained by DMFT, DQMC, and D$\Gamma$A. The gray line 
    is the density of states of the underlying tight-binding model,
    the opaque dotted blue bars show the approximate position
    of the Hubbard bands calculated by Eq.\ (\ref{eq:hubbard-band-position}).}
\end{figure}

The transition also manifests itself in the DMFT quasiparticle renormalization factor $Z$ defined as
\begin{equation}
  \label{eq:qp-weight}
  Z = \bigg[1 - \mathrm{Re}\frac{d\Sigma(\omega)}{d\omega}\Big|_{\omega=0}\bigg]^{-1},
\end{equation}
which decreases from unity in the non-interacting case to zero at the
transition to the insulating state. In~\cfg{fig:qp-renorm}, we illustrate this
behavior for two different temperatures $T=0.33$ and $T=0.1$.  For the latter,
$Z$ vanishes around $U_c\approx9.45$. At the higher temperature, the critical
$U$ is reduced, but its precise estimation is impeded due to thermal
broadening.  In the same plot, we plot a rough estimate for $Z$ from DQMC,
which are estimated as the fraction of the spectral weight located in the
quasiparticle region.  The latter is determined from the DMFT spectral function
for the same $U$ value. For D$\Gamma$A we calculate a momentum-dependent $Z$
which is presented in Fig.~\ref{fig:qp-renorm-u6} below.
\begin{figure}[tb]
  \includegraphics[width=0.5\textwidth]{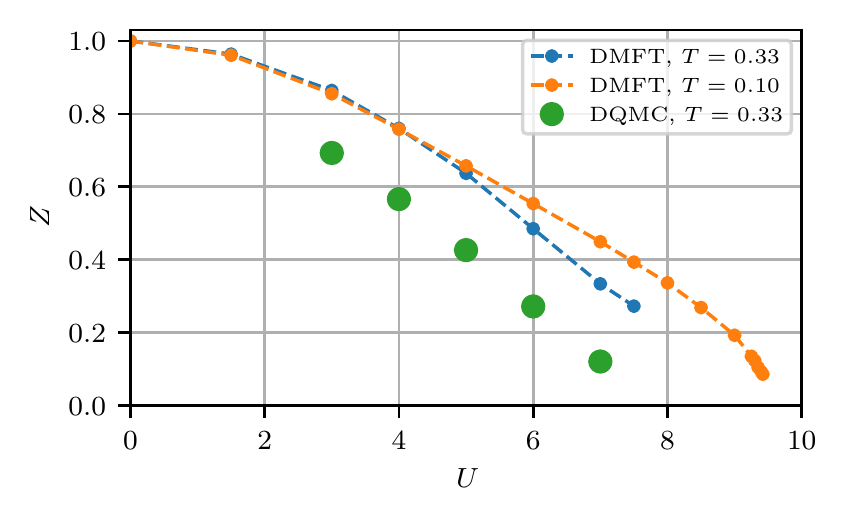}
  \caption{\label{fig:qp-renorm} Quasiparticle renormalization factor $Z$ $[$Eq.~\ref{eq:qp-weight}$]$ based on DMFT and DQMC data as a function of the interaction strength $U$ for the temperatures specified in the legend box.}
\end{figure}

Another characteristic correlation-induced phenomenon is the formation of the Hubbard bands --- two incoherent spectral features separated by $\sim{}U$. As these bands are strongly asymmetric, we estimate their position by
\begin{equation}
  \label{eq:hubbard-band-position}
  \omega_{\text{upper}/\text{lower}} = Z_{\text{DMFT}} \times \omega_{\text{max}/\text{min}} \pm \frac{U}{2}
\end{equation}
where $\omega_{\text{max}/\text{min}}$ are the upper/lower band edge of the tight binding Hamiltonian, 
i.\ e.\ $2\!\!-\!\!\mu_\text{tb}$ and $-4\!\!-\!\!\mu_\text{tb}$.
This heuristic formula interpolates between the non-interacting regime and the
strongly correlated insulting regime where the Hubbard bands are at $\pm\frac{U}{2}$.
It accounts for the fact that the separation of the Hubbard bands is more than
$\pm U$  at small $U$, and describes the position of the Hubbard bands in
\cfg{fig:spec-loc} very well.

\begin{figure}[tb]
  \includegraphics[width=0.5\textwidth]{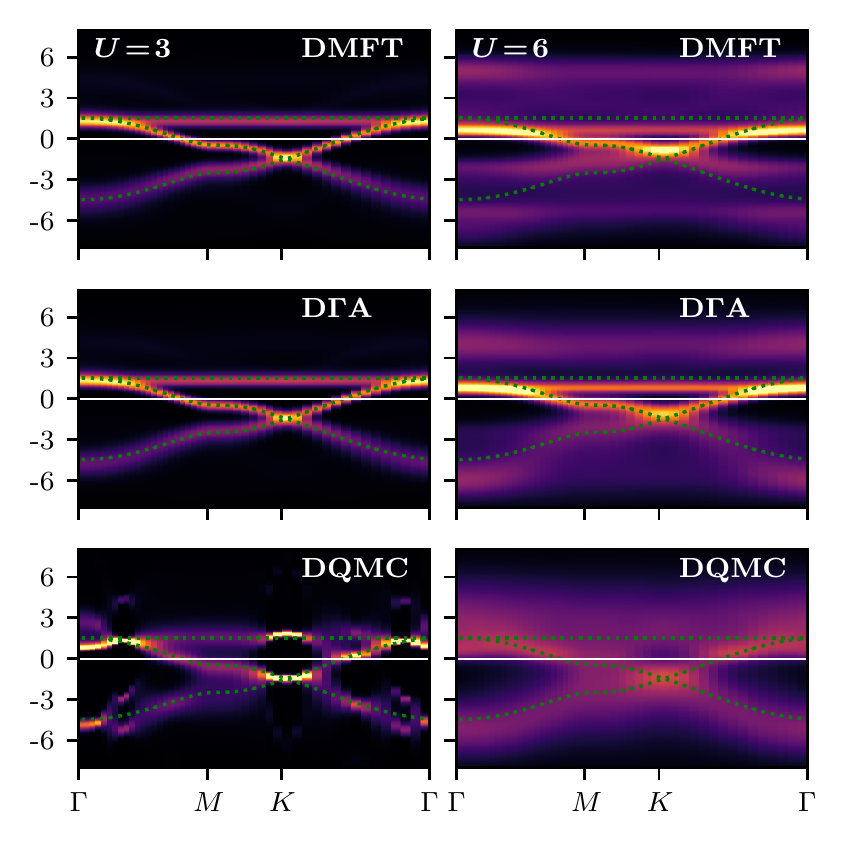}
  \caption{\label{fig:spec-kpath} Spectral functions of DMFT, \dga\ and DQMC on a high-symmetry 
	path through the Brillouin zone. The dotted green lines are the tight-binding
	bands. At weak interaction (left column) there is no sizable renormalization,
	whereas strong interactions considerably change the spectrum (right column).}
\end{figure}

Next, we discuss the momentum-resolved spectral functions in the top panels of \cfg{fig:spec-kpath}. In DMFT, they show a textbook behavior, with a renormalized tight-binding band structure and distinct Hubbard bands. Nonlocal correlations alter this picture: the bottom edge of the quasiparticle band merges with the lower Hubbard band at larger interactions, as can be clearly seen in \dga\ (Fig.~\ref{fig:spec-kpath}, middle panels), and it is even more prominent in DQMC (Fig.~\ref{fig:spec-kpath}, bottom panels). More pronounced in \dga\ spectrum is a waterfall-like structure between the $\Gamma$-point  and $M$/$K$ in the lowest-lying  band below the Fermi energy. Such waterfalls have been observed experimentally in angular resolved photoemission spectra (ARPES) of  cuprate superconductors \cite{PhysRevLett.98.067004}.

We can extract more insights by comparing the DMFT and \dga\ self-energies.
In DMFT, the self-energy is diagonal with respect to  the three basis-lattice sites $j$; all three diagonal elements are even identical and there is no off-site contribution within DMFT.
In \dga, the diagonal elements are almost identical to the DMFT
self-energy and show only a weak dependence on the momentum. The main difference is the presence of sizable off-diagonal elements that are strongly momentum-dependent.

Analysis of matrix-valued quantities computed on the Matsubara axis is not straightforward. 
In the case of the KHM, however, we find that the \dga\ self-energy commutes with the tight-binding Hamiltonian
of \ceq{eq:dispersion-matrix}:
\begin{equation}
  \label{eq:self-energy-commute}
  \big[ \Sigma(\mathbf{k}, i\omega_n), h(\mathbf{k}) \big] = 0.
\end{equation}
For DMFT this is fulfilled trivially, as  $\Sigma$ is $\mathbf{k}$-independent
and proportional to the $3\times3$ unit matrix for the three sites of the unit
cell. In DQMC we do not have direct access to the self-energy, but from the
fact that the lattice Green's function approximately commutes with the
tight-binding Hamiltonian, we can conclude that this holds for the self-energy
as well. This commutation relation implies that the tight-binding bands are
mapped onto interacting quasiparticle bands with associated Hubbard bands, but
they are not mixed by nonlocal correlations, because the interaction is local.
The momentum dependence entails that the self-energy in the band basis is,
albeit diagonal, no longer proportional to the unit matrix as in the DMFT. This
leads to a momentum- and band-dependent, but still well-defined, quasiparticle
renormalization factor $Z_\alpha(\mathbf{k})$, where $\alpha$ is the band
index. We show this quantity for $U=6$ and $T=1/3$ in \cfg{fig:qp-renorm-u6}.
Since smaller $Z$ implies stronger renormalization, it explains why the
lowest-lying quasiparticle band merges with the lower Hubbard band at $\Gamma$,
the center of the Brillouin zone.

\begin{figure}[tb]
  \includegraphics[width=0.5\textwidth]{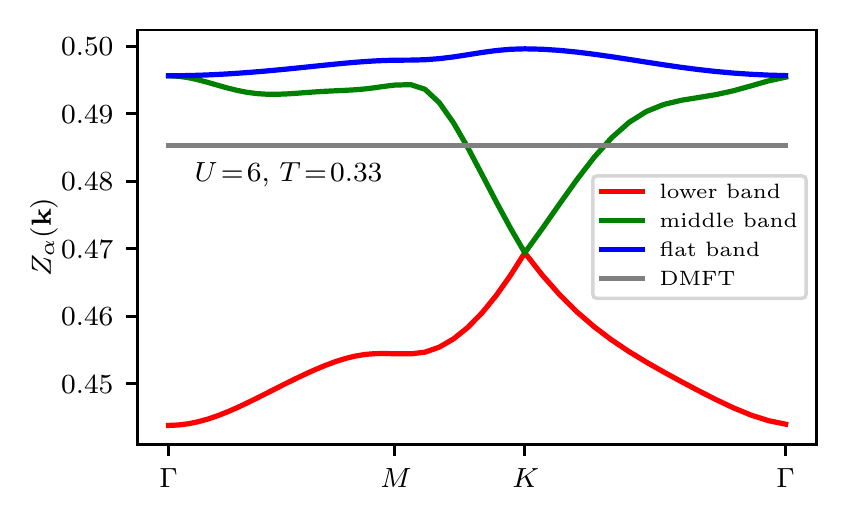}
  \caption{\label{fig:qp-renorm-u6} Momentum- and band-resolved quasiparticle
  renormalization $Z_\alpha(\mathbf{k})$ at $U=6$ and $T=0.33$ obtained by projection
  of the \dga\ self-energy onto the tight-binding eigenbasis.}
\end{figure}

Let us compare these results with the well-studied Hubbard model on a square
lattice. There, antiferromagnetic fluctuations dominate the phase diagram at
all temperatures, leading to an insulating antiferromagnetic state even at
smallest values of the interaction $U$ \cite{schaefer15} in the limit of zero
temperature.  Looking at the magnetic susceptibility of the kagome lattice, we
have to keep in mind that it is defined as
\begin{equation}
  \label{eq:susc-ft}
  \chi_m^{jl}(\mathbf{q}, i\omega_n) = \sum_\mathbf{R} e^{i\mathbf{q}\cdot\mathbf{R}}\hspace{-5pt}
    \int_0^\beta \hspace{-10pt} d\tau \hspace{2pt}e^{i \omega_n \!\tau}
    \langle S^j_z(\mathbf{R}, \tau) S^l_z(\mathbf{0}, 0) \rangle,
\end{equation}
i.e.\ it is a 3$\times$3 matrix for each momentum and frequency. For numerical
and technical reasons our \dga\ calculations are restricted to $\langle S_z^j S_z^l
\rangle$ correlation functions, which yield the standard magnetic
susceptibility. Note that in the absence of magnetic field, diagonal
correlations are equal to one-third of full correlations: $\langle S_z^j S_z^l
\rangle = \frac13\langle \mathbf{S}^j\cdot\mathbf{S}^l \rangle$.
Without symmetry breaking, and as we will see below there are no signs of any long range order, 
spin-off diagonal correlation functions such as $\langle S_z^j S_x^l \rangle$ 
or $\langle S_z^j S_+^l\rangle$ vanish \footnote{Due to the topology of the lattice
with three sites in the unit cell and a triangular Bravais lattice, 
one might expect e.g.\ 120$^{\circ}$ orientations of the spins on neighboring sites. 
However, the $-$120$^{\circ}$ orientation is symmetrically equivalent, 
so that a prevalence toward such spin-orientation
translates to a negative (albeit incomplete) $\left< S_z^j S_z^l \right>$ correlation.}.

For a quantitative analysis, we resort to the eigenvalues of $\chi_m^{jl}$.
Since the susceptibility matrix commutes with the tight-binding Hamiltonian of \ceq{eq:dispersion-matrix},
\begin{equation}
  \label{eq:susc-commute}
  \big[\chi_m(\mathbf{k}, i\omega_n), h(\mathbf{k})\big] = 0,
\end{equation}
we can obtain the three eigenvalues by a projection onto the eigenbasis
of the tight-binding Hamiltonian and thus associate the eigenvalues
with the respective tight-binding bands. In \cfg{fig:chi-ev-bz} we show the (projected) eigenvalues of the zeroth Matsubara frequency 
on the $\mathbf{q}$-plane for $U=3$ at high and low temperature.

\begin{figure}[tb]
  \includegraphics[width=0.5\textwidth]{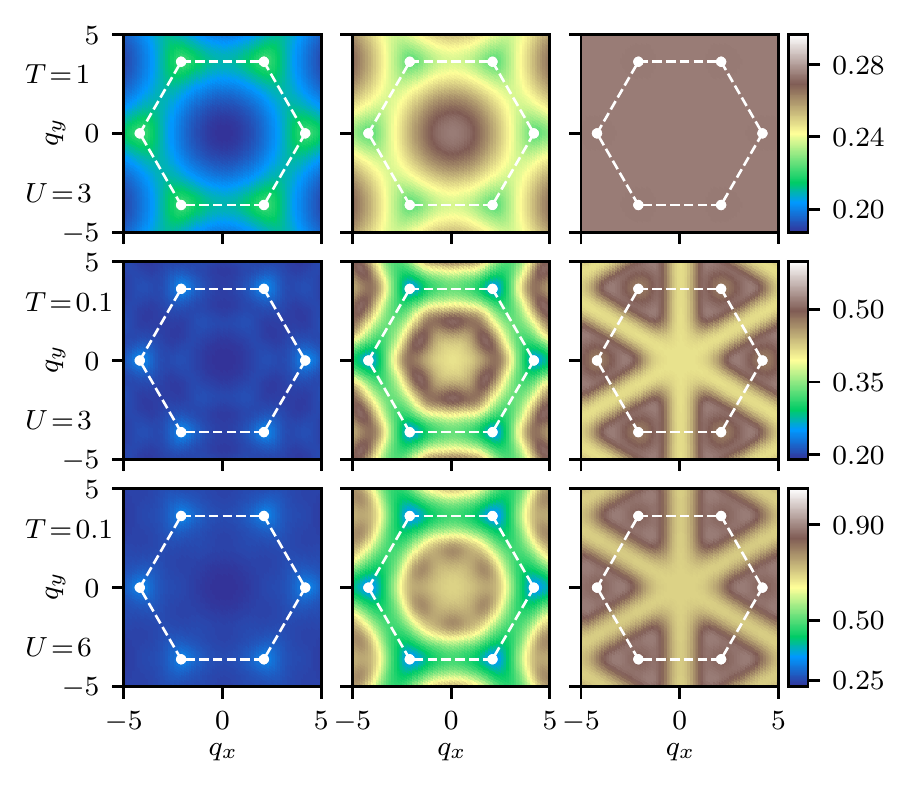}\vspace{-2em}
  \caption{\label{fig:chi-ev-bz} Eigenvalues of the static magnetic susceptibility
    in the first BZ
    for $U=3$ at $T=1$ (upper row) and $T=0.1$ (middle row),
    and $U=6$, $T=0.1$ (lower row).
    There are three eigenvalues corresponding to the three bands in
    \cfg{fig:overview} (c): the left column shows the projection of $\chi$ 
    on the lowest-energy eigenstate [energies below the Dirac point in \cfg{fig:overview} (c)];
    the middle column that of the part of the bandstructure
    above the Dirac point and below the flat band; finally the right column
    corresponds to the flat-band states.}
\end{figure}

\begin{figure}[tb]
  \includegraphics[width=0.5\textwidth]{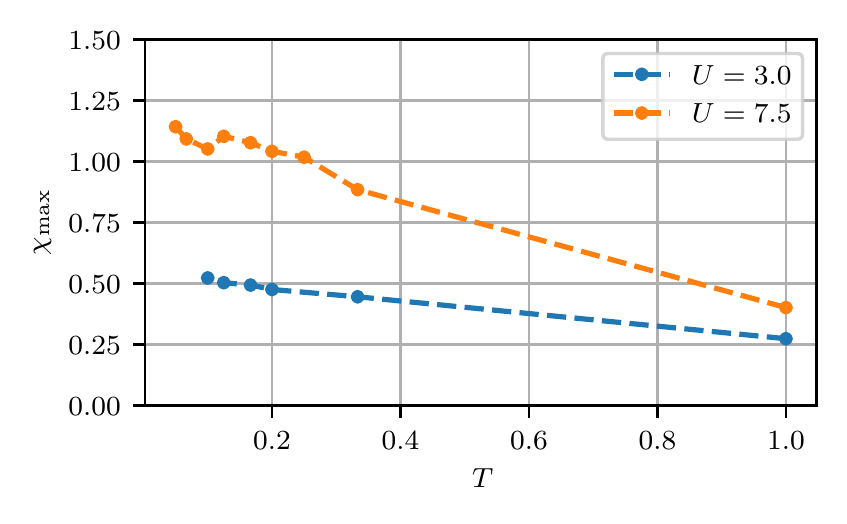}
  \caption{\label{fig:susc-u-temp}Temperature dependence of the maximal magnetic susceptibility $\chi_{\text{max}}$ with respect to the momenta and the three eigenvectors (three panels in  Fig.~\ref{fig:chi-ev-bz}) at $\omega=0$ for the weakly ($U=3$) and moderately ($U=7.5$) correlated regimes.}
\end{figure}

Notably, the projection of the susceptibility on flat-band eigenstates
is also flat at high temperature, but develops an inconspicuous structure at lower temperatures. Most importantly, Fig.~\ref{fig:susc-u-temp} reveals that the maximal value of the susceptibility does not increase significantly as the temperature is lowered. Therefore, in sharp contrast to the square lattice, there is no visible tendency towards magnetic ordering. Instead, the flat structure of the magnetic susceptibility indicates short-ranged spin fluctuations.

\subsection{Structure factors}\label{sec:sq}
Structure factors are on the one hand inherently connected with magnetic susceptibilities, and on the other hand can be addressed experimentally e.g.\ by neutron spectroscopy. Following work on the Heisenberg kagome model~\cite{punk14,sherman18}, we calculate the equal-time structure factor $S_0(\mathbf{q})$ given by
\begin{equation}
  \label{eq:sf-eqt}
  S_0(\mathbf{q}) = \sum_{j,l} e^{i\mathbf{q}\cdot(\mathbf{r}_j - \mathbf{r}_l)}
  \sum_{\omega_n} \chi_{jl}(\mathbf{q}, i\omega_n)
\end{equation}
and the dynamical structure factor $S(\mathbf{q}, \omega)$ given by
\begin{equation}
  \label{eq:sf-dyn}
  S(\mathbf{q}, \omega) = \frac{
    -\mathrm{Im} \sum_{j,l} e^{i\mathbf{q}\cdot(\mathbf{r}_j - \mathbf{r}_l)}
      \chi_{jl}(\mathbf{q}, \omega + i0^+) / \pi
    }{1-e^{-\beta\omega}}.
\end{equation}
In Eqs.~(\ref{eq:sf-eqt}) and (\ref{eq:sf-dyn}), $\mathbf{r}_j$ denotes
the position of the $j$-th atom in the unit cell. 
Since the shortest distance between two sites ($\frac12$) is twice smaller than the unit cell constant (1), the structure factor is periodic in the extended Brillouin zone (Fig.~\ref{fig:overview}, b). 
The analytic continuation of Matsubara frequencies or imaginary time
to real frequencies is performed using ana\_cont~\cite{Geffroy2019, kaufmannGithub} employing the maximum entropy method~\cite{JarrellGubernatis96}.
As a technical remark, we note that in \dga\ we can analytically
continue the projected eigenvalues of the susceptibility matrix individually, and then go back to the sublattice space, where the summation in \ceq{eq:sf-dyn} is carried out. In the following we present our results for the extended BZ of  \cfg{fig:overview} (b).  Upfolding the three eigenvectors of the first BZ  (e.g.\ for the magnetic susceptibility in  Fig.~\ref{fig:chi-ev-bz})  yields a single eigenvector for the larger (extended) BZ.

\begin{figure}[t]
  \includegraphics{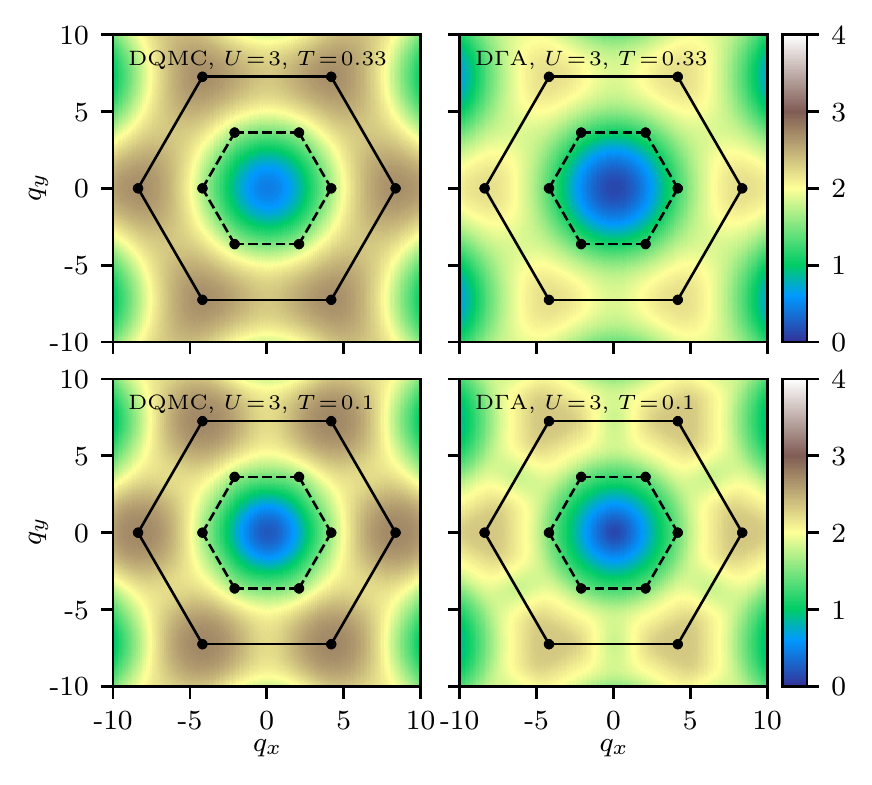}
  \caption{\label{fig:sf-eqt-u3} Equal-time structure factors $S_0(\mathbf{q})$  $[$Eq.~(\ref{eq:sf-eqt})$]$ 
    in the weakly correlated regime ($U=3$) as a function of temperature 
    (top: $T=0.33$, bottom: $T=0.1$). 
    Left: DQMC, right: D$\Gamma$A. 
    The dashed (solid) hexagons denote the boundary of the (extended) Brillouin zones.}
\end{figure}

\begin{table}[b]
\caption{\label{tab:rvecs} 
  Four shortest vectors $\mathbf{R}$ in the kagome lattice, 
  connecting nearest neighbors ($\mathbf{R}_1$), 
  second-neighbors ($\mathbf{R}_2$), and third-neighbors ($\mathbf{R}_{3a}$) 
  and ($\mathbf{R}_{3b}$). Vector components and distances are given in units of the lattice constant.
  (For a visualization of these vectors, see Fig.\ \ref{fig:sf-eqt-latt-u3}.)}
\begin{ruledtabular}
\begin{tabular}{lccl}
notation & $|\mathbf{R}|$ & multiplicity & $\mathbf{R}\equiv\left(R_x, R_y\right)$ \\ \hline
$\mathbf{R}_1$ & $\frac12$ & 4 & $\pm\left(\frac12; 0\right)$, $\pm\left(\frac14; \frac{\sqrt{3}}{4}\right)$ \\
$\mathbf{R}_2$ & $\frac{\sqrt{3}}{2}$ & 4 & $\pm\left(0; \frac{\sqrt{3}}{2}\right)$, $\pm\left(-\frac34; \frac{\sqrt{3}}{4}\right)$ \\
$\mathbf{R}_{3a}$ & 1 & 4 & $\pm\left(1; 0\right)$, $\pm\left(\frac12; \frac{\sqrt{3}}{2}\right)$ \\
$\mathbf{R}_{3b}$ & 1 & 2 & $\pm\left(-\frac12; \frac{\sqrt{3}}{2}\right)$ \\
\end{tabular}
\end{ruledtabular}
\end{table}

\begin{figure}[t]
  \includegraphics{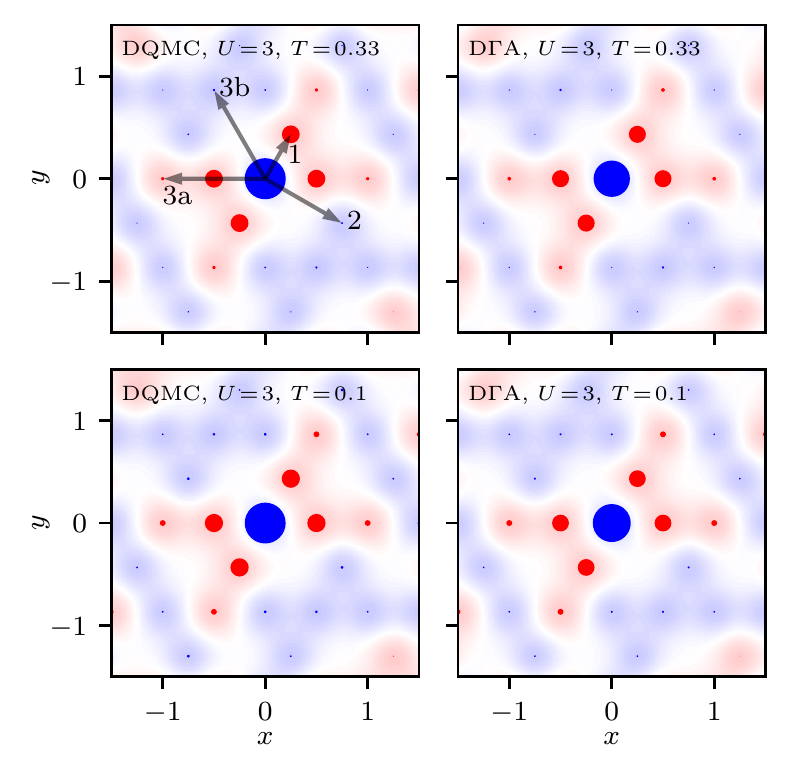}
  \caption{\label{fig:sf-eqt-latt-u3} Equal-time susceptibility $\chi\left(\mathbf{R}\equiv\left(x, y\right)\right)$ 
    in real space for the weakly correlated regime ($U=3$). 
    Units along the $x$ and $y$ axes are unit cell constants. 
    The area of a circle reflects the absolute value of the respective term, 
    blue color denotes positive (ferromagnetic) correlations, 
    red color denotes negative (antiferromagnetic) correlations. 
    Since the magnitude of the long-range correlations is very small,
    we add a background shading to indicate the sign. 
    Note that weak long-range correlations are antiferromagnetic for 
    $\mathbf{R}_{3a}$ and ferromagnetic for $\mathbf{R}_{2}$ 
    and $\mathbf{R}_{3b}$ (see Table~\ref{tab:rvecs} for the notation of intersite vectors).}
\end{figure}
Fig.~\ref{fig:sf-eqt-u3} shows a smooth distribution of the equal-time structure factor 
over the extended Brillouin zone, and a weak temperature dependence. 
In all plots, $S_0(\mathbf{q})$ grows as we move from the center towards 
the boundary of the extended Brillouin zone, 
and forms round maxima at its corners [the K-points in \cfg{fig:overview} (b)]. 
To get a deeper insight into magnetic correlations, 
we return to real space and plot the equal-time susceptibility $\chi$ 
as a function of real-space vector $\mathbf{R}$ connecting two sites. 
As expected for a strongly frustrated model, the resulting pattern 
in Fig.~\ref{fig:sf-eqt-latt-u3} is dominated by two contributions: 
the on-site contribution, which is trivially positive and yields 
a momentum-independent shift, and sizable negative, 
i.\ e.\ antiferromagnetic, correlations between the nearest neighbors ($\mathbf{R}=\mathbf{R}_1$).
As we show in Appendix \ref{app:corr}, the latter are
largely responsible for the maxima at the K-points.

More intriguing is the pattern formed by weak longer-range correlations. 
In particular, by doubling the four $\mathbf{R}_1$, 
we obtain the $\mathbf{R}_{3a}$ vectors (Table~\ref{tab:rvecs}) 
that point to four (out of six) third-neighbors on the kagome lattice. 
A key observation is that the respective correlations are also antiferromagnetic 
(red circles in Fig.~\ref{fig:sf-eqt-latt-u3}). 
In contrast, the second-neighbor correlations ($\mathbf{R}_2$) 
and the remaining two third-neighbor correlations ($\mathbf{R}_{3b}$) 
are ferromagnetic (blue circles in Fig.~\ref{fig:sf-eqt-latt-u3}). 
In momentum space, shown in Fig.~\ref{fig:sf-eqt-u3} this has the 
following effect: strong negative correlations at $\mathbf{R}_1$ 
create the peaks at the K-points and positive correlations at $\mathbf{R}_2$
further increase them.
The negative correlations at $\mathbf{R}_{3a}$ overcompensate
the positive ones at $\mathbf{R}_{3b}$ and thus reduce the structure
factor at the M-points. 
Altogether this leads to well-separated peaks at the K-points.

\begin{figure}[t] 
  \includegraphics{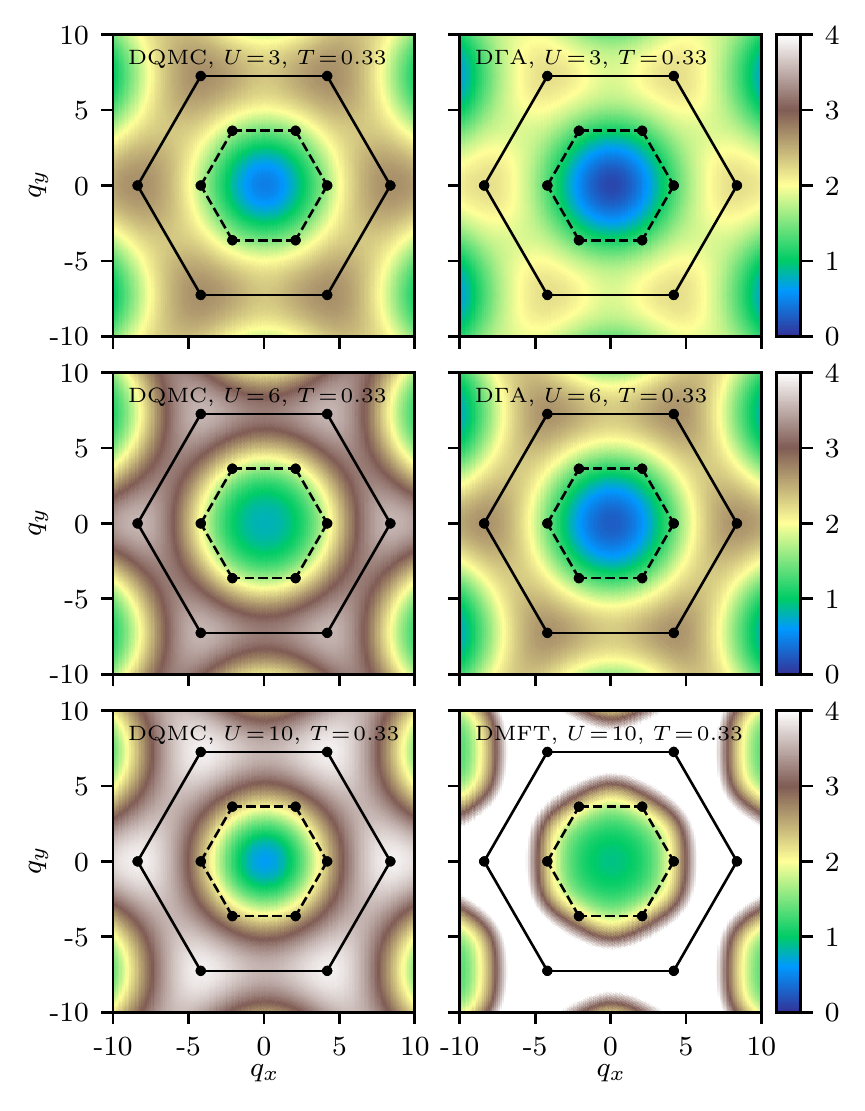}
  \caption{\label{fig:sf-eqt-b3}
Equal-time structure factors $S_0(\mathbf{q})$ $[$Eq.~(\ref{eq:sf-eqt})$]$ as a
function of $U$ (top to bottom) for DQMC (left), \dga (top right, middle
right), and DMFT (bottom right) at $T=0.33$.  We used the DMFT susceptibility
for $U=10$ to calculate the structure factor, as the respective \dga\
calculation did not converge.}
\end{figure}
\begin{figure}[t]
  \includegraphics{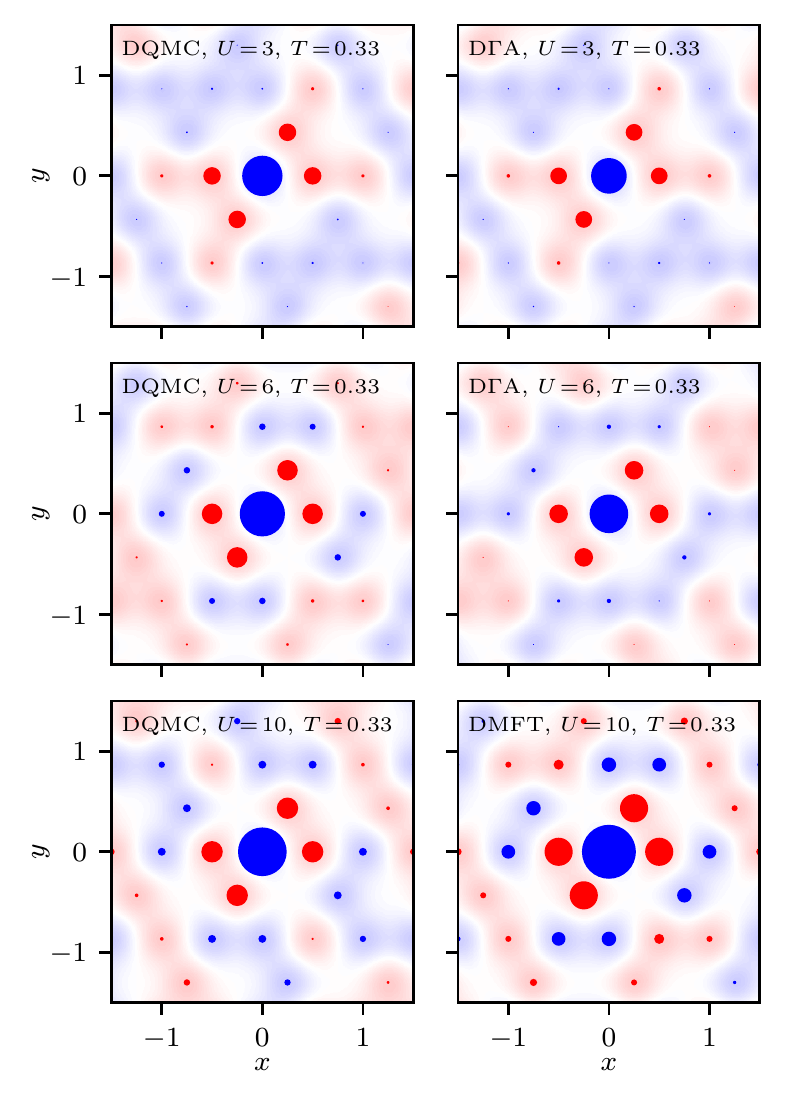}
  \caption{\label{fig:sf-eqt-latt-b3} Equal-time
susceptibility $\chi\left(\mathbf{R}\equiv\left(x, y\right)\right)$ as a
function of $U$ (top: $U=6$, bottom: $U=10$) as calculated by DQMC (left), \dga
(top right) and DMFT (bottom right) at $T=0.33$. The area of a circle reflects
the absolute value of the respective term, blue color denotes positive
(ferromagnetic) correlations, red color denotes negative (antiferromagnetic)
correlations.  Note that weak long-range correlations are ferromagnetic for
$\mathbf{R}_{2}$, $\mathbf{R}_{3a}$ and antiferromagnetic for $\mathbf{R}_{3a}$
(see Table~\ref{tab:rvecs} for the notation of intersite vectors).  }
\end{figure}

So far, we discussed the structure factor in the weakly correlated regime
($U$=3).  If we now increase the interaction $U$, we observe an apparent
change, both in momentum (Fig.~\ref{fig:sf-eqt-b3}) and real
(Fig.~\ref{fig:sf-eqt-latt-b3}) space. While $S_0(\mathbf{q})$ is still peaked
at the K-points, the intensity grows over the entire boundary of the extended
Brillouin zone. 
Again the behavior is understood better by looking at the lattice.
Still, antiferromagnetic nearest-neighbor correlations
generate the dominating peaks at the K-points, supported by $\mathbf{R}_2$-correlations.
However, in the shell of third neighbors now the positive correlations
prevail and increase the structure factor at the M-points, i.\ e.\ \emph{between}
the K-points. This means that the peaks become slightly less separated.
More quantitatively, the ratio between the structure factor at the M-point
and K-point at $U=3$ is 0.87 (0.85) in DQMC (\dga), and it increases to
0.90 (0.87) at $U=6$.

Interestingly, our patterns
are in excellent agreement with the equal-time structure factor
computed using numerical linked cluster expansion for the Heisenberg kagome
model~\cite{sherman18}. We will discuss the ramifications in
Sec.~\ref{sec:discussion}.

Since our calculations provide direct access to dynamical quantities, it is
instructive to inspect the energy dependence of the structure factors. In
\cfg{fig:sf-qpath-u3} we plot the dynamical structure factor $S(\mathbf{q},
\omega)$ on a path through the extended Brillouin zone
(Fig.~\ref{fig:overview}) at weak interaction $U=3$ for two different
temperatures. In line with the equal-time structure factor, the dominant weight
is located around the K-point. Additionally, there is a splitting into a low- and
a high-energy mode at the M-point (cf.~Fig.~\ref{fig:overview}, b), although
at high energies it is concealed by thermal broadening in \dga. 

\begin{figure}[tb] \includegraphics[width=0.5\textwidth]{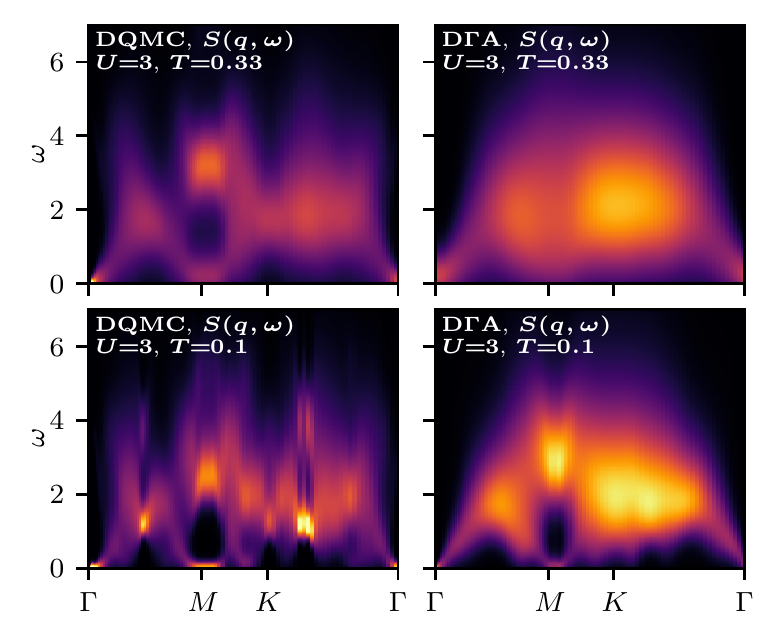}
\caption{\label{fig:sf-qpath-u3} Dynamical structure factor $S(\mathbf{q},
\omega)$ in the weakly correlated regime ($U=3$) at $T=0.33$ (top) and $T=0.1$
(bottom) calculated using DQMC (left) and \dga\ (right) on a path through the
extended Brillouin zone (Fig.~\ref{fig:overview}).} \end{figure}

\begin{figure}[tb]
  \includegraphics[width=0.5\textwidth]{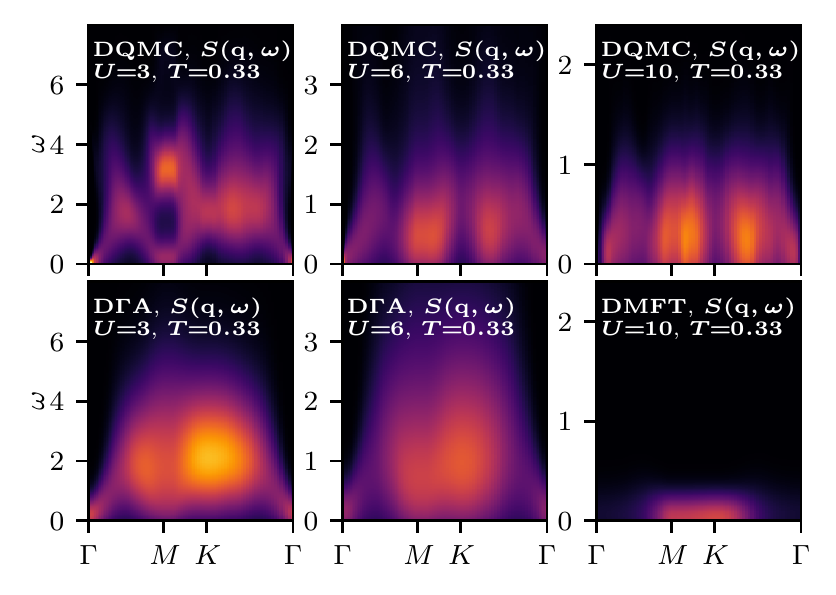}
  \caption{\label{fig:sf-qpath-b3} Dynamical structure factor 
        as a function of interaction strength (left to right)
        as calculated by DQMC (top) and \dga (bottom, except for DMFT in
the right bottom plot) on a path through the extended Brillouin zone
  (Fig.~\ref{fig:overview}) at $T=0.33$. Note that the upper limit of the $\omega$-axis ($y$-axis)
is reduced as $1/U$ upon increasing $U$ to resolve the feature-rich range; 
for large $U$ such a rescaling must hold as the coupling of the Heisenberg model is $J=4t^2/U$.}
\end{figure}

Finally, we present the evolution of the dynamical structure factor as a
function of correlation strength in Fig.~\ref{fig:sf-qpath-b3}. The main effect
is squeezing the frequency spread of the intensity to lower energies
accompanied by a gradual dissipation of spectral features. Note that it was not
possible to converge a self-consistent \dga\ calculation for the $U=10$ case,
and in DMFT the spectral weight is pushed to zero energy.

\section{Discussion}\label{sec:discussion}
Diagrammatic extensions alleviate the main drawback of DMFT --- its restriction
to local correlations. An alternative route to include nonlocal correlations
are cluster extensions of DMFT. In the conceptually simple cellular DMFT
(CMDFT) approach, such clusters are constructed in real space and typically
comprise a small number of sites. Correlations in CDMFT are still purely local,
but the locality spans now the entire cluster. As a result, nonlocal
correlations at the length scale of the cluster are included, while
longer-range fluctuations are still absent. In addition to this sharp cutoff
between the short-range (included) and long-range (omitted) correlations, CDMFT
introduces a spurious disparity between the sites falling within the cluster
and all other sites. We can illustrate this by considering the kagome lattice,
where each site has four equivalent nearest neighbors. In the simplest possible
cluster, a triangle, this equivalence is violated: only two of the four
neighboring sites belong to the cluster. Keeping these intrinsic limitations of
CDMFT in mind, we compare the results of three different CDMFT
studies~\cite{ohashi06, udagawa10, kita13} with our \dga\ and DQMC results.

In three-site CDMFT employing a Hirsch-Fye impurity solver, Ohashi \etal\
reported a first-order transition at the critical $U_c=8.22$~\cite{ohashi06}.
While we were not aiming at a precise estimate for the critical interaction
strength, the three-site CDMFT values is somewhat lower than our DMFT value and
higher than our DQMC estimate with 75 sites (and no DMFT bath). 

In the strongly correlated regime ($U=6.6$), the largest eigenvalue of the
magnetic susceptibility plotted as a function of $\mathbf{q}$, is nearly flat
in CDMFT, with shallow minima along the six $\Gamma$-M lines. Precisely this
behavior is observed in the \dga\ susceptibility (Fig.~\ref{fig:chi-ev-bz}).
Interestingly, Ohashi \etal\ report a drastic change of magnetic correlations
in the insulating phase: they argue that a starlike structure in the structure
factor indicates the onset of 1D antiferromagnetic
correlations~\cite{ohashi06}. We believe that this an artifact of CDMFT. This
is corroborated by the fact that  the spin correlations that we obtain in \dga\
and DQMC for the moderately correlated regime of the Hubbard model are similar
to those of the Heisenberg model~\cite{sherman18}.

The more recent CMDFT study by Udagawa \etal~\cite{udagawa10} employs a
continuous-time auxiliary-field QMC impurity solver and uses, in addition to
triangles, also more extended nine-site clusters. While the density matrix
defined from microscopic states of a cluster is not accessible in the
diagrammatic extensions, our temperature-dependent susceptibility
(Fig.~\ref{fig:susc-u-temp}) agrees with the CDMFT results plotted in Fig. 4(d)
of Ref.~\cite{udagawa10}, except for the lowest temperatures, where we do not
find the downturn of $\chi(T)$ characteristic for antiferromagnetic
correlations or the formation of localized dimers.  While not much is known for
the KHM, we note that in the kagome Heisenberg model, dimer tunneling processes
around the loops comprising eight sites play a pivotal role~\cite{ralko18}. The
absence of such loops in the nine-site clusters used in Ref.~\cite{udagawa10}
may give rise to the formation of static antiferromagnetic dimers, and hence a
suppressed susceptibility.

Finally, Kita \etal\ focus on the behavior of the KHM in a magnetic
field~\cite{kita13}. Nevertheless, it is instructive to discuss their CDMFT
spectral function in zero field, computed for $U=4$ and $U=8$ (Fig.~1 in
Ref.~\cite{kita13}). The former agrees well with our \dga\ results for $U=3$,
except for the substantial broadening in the CDMFT data at low frequencies,
which might be an artifact of our analytic continuation. The $U=8$ case is more
interesting.  Here, in contrast to single-site DMFT and in agreement with \dga,
no distinct lower Hubbard band is formed; instead, a dispersive feature
stemming from the lowest-lying branch of the kagome band structure shows up at
low frequencies. Close to the Fermi level, two narrow, nearly dispersionless
bands form in CDMFT\cite{kita13}. We do not observe such structures in our
$U=6$ calculations, neither in \dga, nor in DQMC: the intensity maxima lie
above the Fermi level. At higher frequencies corresponding to the upper Hubbard
band, CDMFT shows a broad spectral maximum in the vicinity of the
$\Gamma$-point which rapidly decreases for finite momenta. A similar, albeit
less pronounced distribution of the spectral weight is visible in \dga\ results
(Fig.~\ref{fig:spec-kpath}): the intensity at the $\Gamma$-point is maximal.

Structure factors are a direct source of information on the dominant magnetic
correlations and instabilities. A prominent example is the antiferromagnetic
instability of the square-lattice Hubbard model, signaled by the diverging
structure factor at $\mathbf{q}=\left(\pi,\pi\right)$, the propagation vector
of the N\'eel state. In contrast, the structure factors of the KHM in the
metallic regime lack any apparent instabilities, and instead show an intricate
evolution on the frequency/momentum grid. Thus, to get insights into the
magnetic correlations, we compare the behavior of $S(\mathbf{q})$ and
$S(\mathbf{q,\omega})$ with the literature data for the Heisenberg model.

The kagome Heisenberg model features several low-lying states with marginally
different energies. While the debate on the ground state is still not settled,
structure factors recently came into the forefront as a possible fingerprint to
distinguish these states experimentally. A popular strategy is to pick a
certain candidate state and calculate its structure factor using various
mean-field techniques~\cite{messio10, iqbal13, dodds13, punk14, halimeh16,
messio17, halimeh19}. However, in the context of our study, a more appropriate
starting point is the direct simulation of the Heisenberg model on a finite
lattice, followed by the evaluation of structure factors from the spin
correlations.  Regardless of the method used, the resulting $S(\mathbf{q})$ of
the Heisenberg model smoothly evolves from the minimum at $\Gamma$ to the
maximum at the boundary of the extended Brillouin zone. Further details depend
on the computational method: While exact diagonalization on 36-site finite
lattices yields feeble, yet discernible peaks at M~\cite{laeuchli09X,
seman15X}, these features are practically wiped out in density-matrix
renormalization group (DMRG) simulations~\cite{kolley15, zhu19} that are less
prone to finite-size effects. The featureless structure factor indicates that
tendencies to ordering are strongly suppressed, even on a short range. Since
they arise from competing correlations, this balance can be destroyed by small
deviations from the Heisenberg model, such as anisotropies and/or longer-range
exchanges. A common ramification is the appearance of maxima at K or M points
of the extended Brillouin zone, indicative of so-called \cite{messio11} \sqsq\ or \qzero\
antiferromagnetic correlations, respectively.

We are now in the position to compare the equal-time structure factors of the
KHM in Figs.~\ref{fig:sf-eqt-u3} and \ref{fig:sf-eqt-b3} with that of the
Heisenberg model. First, the smooth $\mathbf{q}$-evolution and the minimum at
$\Gamma$ are common for both models. For all studied $U$ and $T$ values, the
maximal intensity is at K, indicating the predominance of \sqsq\ correlations.
This is seemingly at odds with the Heisenberg model, where weak maxima, if any,
are found at M~\cite{laeuchli09X, seman15X}. However, a key difference lies in
the methods: we do calculations at finite temperature. Looking at the
finite-temperature structure factors for the Heisenberg model~\cite{sherman18},
we see a strikingly similar picture: a smooth evolution with maxima at K. This
brings us to one of the main conclusions: at moderate temperatures, the
magnetic correlations of the KHM are similar to those of the Heisenberg model.

While our equal-time structure factors are quantitatively similar in the weakly
and strongly correlated regime (cf.\ Fig.~\ref{fig:sf-eqt-u3} and
Fig.~\ref{fig:sf-eqt-b3}), real-space plots of respective susceptibilities
reveal a subtle change in third-neighbor correlations (cf.\
Fig.~\ref{fig:sf-eqt-latt-u3} and Fig.~\ref{fig:sf-eqt-latt-b3}). 
At weak coupling ($U=3$), Fig.~\ref{fig:sf-eqt-latt-b3} shows 
along the direction of the two Bravais lattice vectors a 
negative($\mathbf{R}_1$)-negative($\mathbf{R}_{3a}$)-positive 
spin correlation function, hinting at tendencies toward a 120$^{\circ}$
spin-orientation. At strong coupling ($U=6$ and $U=10$) the 
third nearest-neighbor ($\mathbf {R}_{3a}$) changes sign. 
In the half-filled one-band Hubbard model with a strong negative 
(antiferromagnetic) preference between nearest neighbors, 
this is arguably the most dramatic change one might expect, 
devoid an actual ordering that is prevented by the frustrated lattice. 
Note that the third nearest neighbors and second nearest neighbors 
are both at a distance of two hopping elements, 
they only differ by their distance in real space because of the geometry.

We attribute
this difference to the correlation-induced onset of \qzero\ magnetic
correlations that compete with dominating \sqsq\ correlations. At the same
time, susceptibilities in the moderately correlated regime
(Fig.~\ref{fig:sf-eqt-latt-b3}) are qualitatively similar to those in the
Heisenberg model~\cite{sherman18}. Therefore, we conclude that $\mathbf{q}=0$
correlations develop already in the moderately correlated regime, i.e.\ in the
metallic phase. This nontrivial result provides a key to distinguish between
the weakly and moderately correlated regimes in real materials: The former
features predominantly \sqsq\ correlations, while in the latter additional
\qzero\ correlations become manifest.

Next, we discuss the features of the dynamical structure factor $S(\mathbf{q},
\omega$). While details of the plots are prone to uncertainties of the analytic
continuation, we comment on one salient feature: the difference between the
frequency dependencies at K and M. The highest spectral density is associated
with K (consistent with the maxima in the equal-time structure factor), but is
it shifted to higher frequencies as compared to M.  Interestingly, the same
structure is found in the structure factor of a $Z_2$ spin liquid with a
moderate spinon-vison interaction~\cite{punk14}.  Our work should motivate
further studies to clarify whether metallic kagome magnets can serve as a
playground for topological vison excitations, which have been suggested
in~\cite{punk14}.

Finally, we put our results in the context of ongoing experimental activities
on metallic kagome magnets. While we computed the quantities that can be
measured by inelastic neutron scattering, several aspects impede a direct
comparision.  First, all so far discovered metallic kagome materials are
multi-orbital systems. A simplified effective one-orbital description is
generally possible, but the mapping scheme depends on the specifics of a
particular material and has to be adjusted accordingly. Second, a kagome-like
arrangement of magnetic atoms in the crystal structure does not guarantee the
applicability of the KHM: coupling beyond nearest neighbors as well as interplane
couplings can play a significant role. This is the case for Mn$_3$Sn, where
neutron scattering experiments reveal the relevance of multiple magnetic
exchanges~\cite{kang19}.  Bilayer kagome systems Fe$_3$Sn$_2$~\cite{tanaka20}
and Co$_3$Sn$_2$S$_2$~\cite{liu21} that entail a sizable interlayer coupling
fall in the same category.  Also the band filling, whose estimation in a real
material is per se challenging, can deviate from the case in point: KHM at
strict half-filling.  All in all, we believe that presently the most promising
case is FeSn, whose band structure (Fig.~4 in Ref.~\cite{kang19}) bears
apparent similarities to the half-filled tight-binding kagome model. We are
looking forward to future inelastic neutron scattering experiments (announced
in Ref.~\cite{sales19}) that can be compared with our structure factors and
dynamical susceptibilities.

\section{Conclusion}\label{sec:conclusion}
We studied the phase diagram and the magnetic structure factor of the kagome
Hubbard model, focusing on the weakly and moderately correlated regime relevant
for the growing family of real materials.  To this end, we employed three
complementary methods: DMFT, \dga\ and DQMC. We observe neither tendencies
towards magnetic ordering of any kind, nor fingerprints of singlet formation.
To provide solid reference data for inelastic neutron scattering experiments on
candidate materials, we calculate dynamical as well as equal-time structure
factors and susceptibilities, for a wide range of the interaction parameters
$U$ and at different temperatures.  By comparing our results with the
literature data for the Heisenberg model, we conclude that the Mott transition
is not accompanied by a sensible alteration of magnetic correlations: the major
change happens already in the metallic phase, where the magnetic coupling to
third-nearest neighbors changes sign. We argue that this change gives a key to
estimate $U$, and hence the proximity to a metal-to-insulator transition, in
real materials.

\begin{acknowledgements}
We are grateful to Daniel Hirschmeier and Alexander Lichtenstein 
for inspiring discussions and their help in the early stages of this project; 
furthermore, we thank them and Andrey Lehmann 
for sharing preliminary results of their dual fermion calculations.
We acknowledge fruitful discussions with Evgeny Stepanov, 
Joseph Checkelsky, Linda Ye, Shiang Fang, Satoshi Nishimoto, 
and Johannes Richter.  J.\ K.\ further thanks Malte
R\"osner, Clio Agrapidis and Lukas Rammelm\"uller for useful comments.
Calculations have been done on the Vienna Scientific Cluster~(VSC) and the
computational facilities of the Leibniz IFW Dresden.  We thank U.~Nitzsche for
technical assistance.  Plots were made using the matplotlib~\cite{hunter07}
plotting library for python.  J.\ K.\ and K.\ S.\ thank the  Leibniz IFW
Dresden and UC Davis, respectively, for hospitality during their stay.  O.\,J.\
was supported by the Leibniz Association through the Leibniz Competition; J.\
K.\ and K.\ H.\ by the Austrian science fund (FWF) through projects P~32044 and
P~30997; K.\ S.\ by the Marshall Plan Scholarships Program; the work of R.\ T.\
S.\ was supported by the grant DE--SC0014671 funded by the U.S. Department of
Energy, Office of Science.
\end{acknowledgements}

\appendix*
\section{Fingerprint of real-space correlations}
\label{app:corr}
\begin{figure*}
    \centering
    \includegraphics{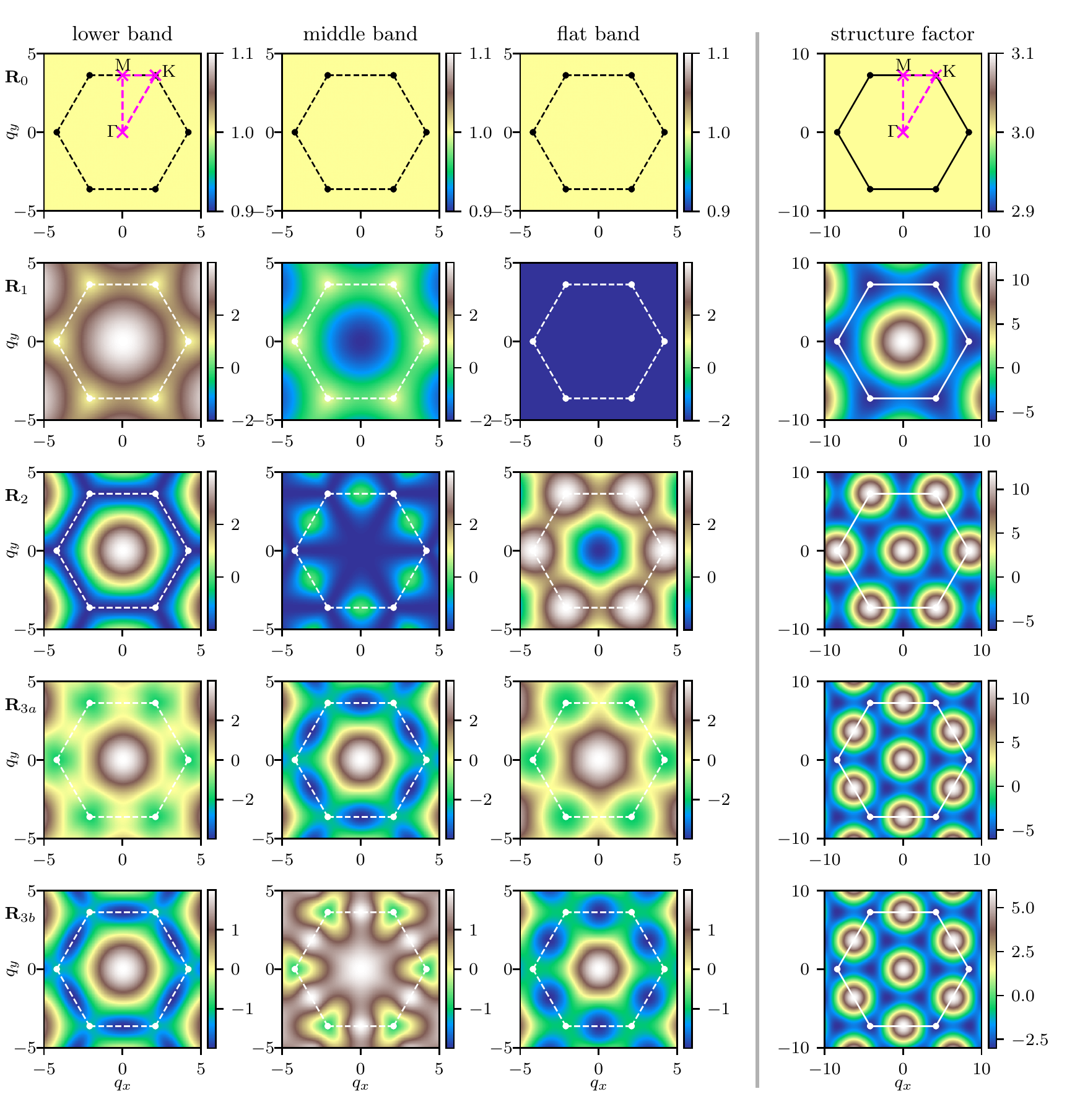}
    \caption{\label{fig:app01} How certain real-space correlations influence
    the magnetic susceptibility and structure factor. 
    In the first three columns we show the projection of the
    magnetic susceptibility on the lower-band, middle-band and flat-band
    eigenstates of the tight-binding Hamiltonian, respectively.
    The right-most column shows the corresponding structure factor $S(\mathbf{q})$ on 
    the extended Brillouin zone.
    The rows correspond to various correlations in real space: 
    Thus in the first row we show momentum-space correlations arising
    from on-site correlations. 
    The second row corresponds to nearest-neighbor correlations ($\mathbf{R}_1$).
    The third, fourth and fifth row correspond to $\mathbf{R}_2$, $\mathbf{R}_{3a}$
    and $\mathbf{R}_{3b}$ correlations.
    }
    \vspace{1cm}
\end{figure*}

For a better understanding of how spin-spin correlations between certain points in 
the lattice affect the susceptibility or structure factor in momentum space,
it is helpful to study the connection analytically.
The basis for this is Eq.\ (\ref{eq:susc-ft}), where we set 
$\langle S^j_z(\mathbf{R}, \tau) S^l_z(\mathbf{0}, 0) \rangle = 1$
for a certain vector $\mathbf{R}$ and all vectors that are related by
symmetry transformations.
Fig.\ \ref{fig:app01} shows, row by row, 
how correlations to a certain neighbor $\mathbf{R}$ 
and its symmetrically related counterparts reflect in $\mathbf{k}$-space.
The first three rows are the projections to
tight-binding eigenstates, and the fourth row is the structure factor. 

Unsurprisingly, on-site correlations yield just a constant contribution.
$\mathbf{R}_1$- and $\mathbf{R}_2$-correlations lead to peaks at the $K$-point
in the extended Brillouin zone, whereas $\mathbf{R}_3$-correlations enhance the
$M$-point (last two rows in Fig.\ \ref{fig:app01}).

It  is important to note that, for this analysis, 
we always consider positive correlations (of unit magnitude) between neighboring sites 
at the indicated distance, negative ones just change the sign. 
For the nearest-neighbor correlations ($\mathbf{R}_1$) this means, e.g., 
that we get negative peaks at the K-points, whereas the actual 
correlations at $\mathbf{R}_1$ are negative yielding positiv peaks around the K-points.
Furthermore let us note that the number of neighbors at $\mathbf{R}_{3a}$ is
twice as large as the number of neighbors at $\mathbf{R}_{3b}$. Therefore
also their influence on the structure factor is twice as large.

\newpage

\bibliography{bibliography}
 
\vfill\eject

\end{document}